\documentclass[useAMS,usenatbib]{mn2e}
\usepackage[percent]{overpic}
\usepackage{times}
\usepackage{graphicx}
\usepackage{amsmath}
\usepackage{mathabx}
\usepackage{subfigure}
\usepackage{natbib}
\usepackage{txfonts}
\usepackage{journals}
\usepackage{xcolor}
\usepackage[utf8]{inputenc}
\usepackage[overload]{empheq}

\newcommand{\kms}{$\rm km\,s^{-1}$}

\newcommand{\ks}{$K_{\rm S}$}
\newcommand{\aks}{$A_{K_{\rm S}}$}

\title[Distances of molecular clouds]
{The distances to molecular clouds in the fourth Galactic quadrant} 

\author[B.Q. Chen et al.]
{Bingqiu Chen,$^{1}$\thanks{E-mail:
bchen@ynu.edu.cn (BQC); shuwang@nao.cas.cn (SW).}
Shu Wang,$^2$\footnotemark[1]
Ligang Hou,$^{2}$
Yihong Yang,$^{1,3}$
Zhiwen Li,$^{1,3}$
He Zhao,$^{4,5}$ 
\newauthor
and  Biwei Jiang$^4$
\\
$^{1}$South-Western Institute for Astronomy Research, Yunnan University, Kunming, 650500, P.\,R.\,China\\
$^{2}$National Astronomical Observatories, Chinese Academy of Sciences, Beijing 100101, P.\,R.\,China\\
$^{3}$School of Physics and Astronomy, Yunnan University, Kunming, 650500, P.\,R.\,China\\
$^{4}$Department of Astronomy, Beijing Normal University, Beijing, 100875, P.\,R.\,China\\
$^{5}$University Cote d'Azur, Observatory of the Cote d’Azur, CNRS, Lagrange Laboratory, Observatory Bd, CS 34229, F-06304 Nice cedex 4, France \\
 }

\begin{document}

\date{Accepted ???. Received ???; in original form ???}

\pagerange{\pageref{firstpage}--\pageref{lastpage}} \pubyear{2020}
\maketitle
\label{firstpage}

\begin{abstract}
Distance measurements to molecular clouds are essential and important. We present directly measured distances to 169 molecular clouds in the fourth quadrant of the Milky Way. Based on the near-infrared photometry from the Two Micron All Sky Survey and the Vista Variables in the Via Lactea Survey, we select red clump stars in the overlapping directions of the individual molecular clouds and infer the bin averaged extinction values and distances to these stars. We track the extinction versus distance profiles of the sightlines toward the clouds and fit them with Gaussian dust distribution models to find the distances to the clouds. We have obtained distances to 169 molecular clouds selected from Rice et al. The clouds range in distances between 2 and 11\,kpc from the Sun. The typical internal uncertainties in the distances are less than 5\,per\,cent and the systematic uncertainty is about 7\,per\,cent. The catalogue presented in this work is one of the largest homogeneous catalogues of distant molecular clouds with the direct measurement of distances. Based on the catalogue, we have tested different spiral arm models from the literature. 
\end{abstract}


\section{Introduction}

Distance is the fundamental parameter to estimate all other 
physical properties like the size and mass of molecular cloud.
Estimating the distance to molecular cloud is a tough task and
astronomers have explored several different techniques. 

A first method is to derive the distance kinematically,
which convert the radial velocity of a cloud to a distance by assuming that the cloud
follows the Galactic rotation curve (e.g., \citealt{Roman2009, Miville2017}). 
However, the kinematic distances are problematic due to the large uncertainty of the rotation
curve and the influence of the peculiar velocities and the noncircular
motions. Furthermore, a well-known geometric ambiguity exists
for the kinematic distance of cloud in the inner Galaxy, where
one velocity can be related to two distances.
A second method is to obtain the distance of a given cloud by identifying its
associated objects having the same distance as the cloud, 
such as the OB stars and young stellar objects, whose distances can be estimated (e.g. \citealt{Gregorio2008}). 
However, this method is only applied to individual clouds of interest. 
A third method is to estimate the distance from the extinction of starlight, i.e., the extinction distance. 
As the density of the dust grains in the molecular cloud is much higher than that in diffuse medium, 
one can expect a sharp increase of stellar extinction at the location of the cloud. 
Thus the distance to the cloud can be obtained by finding the position where the extinction increases sharply. 

The extinction method is directly measured and robust. 
However, it relies on the accurate estimates of distances and extinction values of a large number of stars.  
Thanks to a number of large-scale astrometric, photometric and spectroscopic surveys,
we can obtain accurate values of distance and dust extinction 
for tens of millions of individual stars (e.g., \citealt{Chen2014, Green2015}).
Thus the estimation of precise extinction distances to large samples of molecular clouds has become possible.  
\citet{Schlafly2014} obtain distances to dozens of molecular clouds selected from the literature by the three-dimensional (3D) extinction mapping method based on PanSTARRS-1 data.
\citet{Zucker2019} improve their work by combining the optical and near-IR photometry with the Gaia Data Release 2 (Gaia DR2; \citealt{Gaia2018}) parallaxes \citep{Lindegren2018}.
\citet{Yan2019} derive extinction distances to 11 molecular clouds in the third Galactic quadrant.
\citet{Chen2020} present accurate distance determinations to a catalogue of 567 molecular clouds based on estimates of colour excesses and distances of stars presented in \citet{Chen2019}.
\citet{Zhao2018, Zhao2020} estimate the extinction distances to 33 supernova remnants with the help of their association with molecular clouds.

All the above works calculate only the extinction distances to the nearby molecular clouds locating within $\sim$4\,kpc from the Sun. 
These works are based on the optical extinction which becomes very large at far distances and they adopt the
trigonometric parallax, photometric or spectroscopic distances of stars
that suffer large uncertainties at far distances. Thus they are not able to determine the distances to far clouds. 
At far distances, the standard candle red clump star (RC) serves as a good tracers to 
infer the extinction distances of molecular clouds. RCs have almost constant luminosity and color.
They are bright in the near-infrared (IR) bands and can trace the extinction profile at far distances 
\citep{Gao2009, Guver2010, Gonzalez2012}. 
\citet{Shan2018, Shan2019} use RCs to map the 3D extinction distribution of
the sightlines toward the supernova remnants in the
first and fourth quadrants of the Galaxy
and estimate distances of them with extinction values from the literature.
\citet{Wang2020} use RCs to estimate extinction distances of 63 molecular clouds 
 and infer them to the distances of supernova remnants.

In this work, we select RCs from the near-IR colour-magnitude diagram (CMD) to determine the 
distance to a large sample of molecular clouds in the fourth Galactic quadrant based on the data from the 
Two Micron All Sky Survey (2MASS; \citealt{Skrutskie2006}) and the 
Vista Variables in the Via Lactea Survey (VVV; \citealt{Minniti2010}).
We select molecular clouds from \citet{Rice2016}, who present a catalogue of 1,064 massive molecular 
clouds throughout the Galactic plane from the CO survey of \citet{Dame2001}. 
We have estimated robust, directly-measured distances of 169 molecular clouds. 
The most distant cloud is located at a distance larger than 11.5\,kpc from the Sun.

The paper is structured as follows. In Section\,2, we describe the data. In Section\,3 we 
introduce our method for determining the distances to the molecular clouds. We present our 
results in Section\,4, and discuss them in Section\,5. We summarize in Section\,6.

\section{Data}

\begin{figure*}
    \centering
  \includegraphics[width=0.99\textwidth]{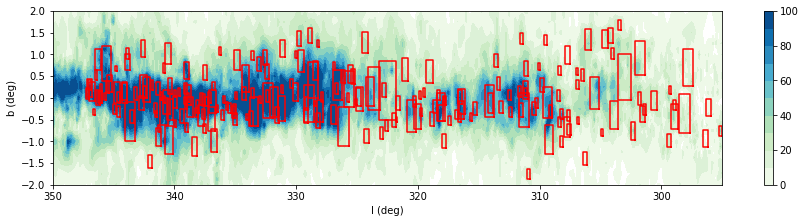}
  \caption{Distributions of the selected molecular clouds in the Galactic coordinates. 
  The background image illustrates the distributions of integrated CO emission in 
  the fourth quadrant from \citet{Dame2001}. Red rectangles mark the $l$ and $b$ boundaries 
  of the individual molecular clouds adopted in the current work.}
  \label{catalog}
\end{figure*}

Our work is based on near-IR photometry from 2MASS and VVV. 
2MASS was conducted with two 1.3\,m telescopes to provide full coverage of the sky. 
It has three near-IR bands, $J, ~H$ and \ks, centred at 1.25, 1.65, and 2.16\,$\mu$m, respectively. 
The 2MASS Point Source Catalog (PSC; \citealt{Skrutskie2006}) contains near-IR photometry for 
over 500 million objects. The systematic uncertainties of 2MASS photometric measurements are 
estimated to be smaller than 0.03\,mag. The 10$\sigma$ \ks\ limiting magnitude is approximately 14.3\,mag. 
To trace the distance of far molecular clouds, we combined the 2MASS data with the in-depth near-IR VVV 
data. The VVV survey were carried out with the VISTA 4.1\,m telescope. 
The survey took images in five bands, $Z, ~Y, ~J, ~H$ and \ks. The total survey area of VVV is 
560\,deg$^2$, including a region of the Galactic plane of 295\degr $<~ l ~<$ 350\degr\ to $-$2\degr $<~ b ~<$2\degr and 
the Bulge of $-$10\degr $<~ l ~< $10\degr\ and $-$10\degr $<~ b ~<$ 5\degr. 
In the current work, we adopt the catalogue  `vvvPsfDophotZYJHKsSource'  from the Data Release 4 of Images and Source 
Lists from VVV\footnote{https://www.eso.org/sci/publications/announcements/sciann17009.html}.
The VVV \ks\ saturation limit ranges between 10 -- 12\,mag, and the photometric limit is typically 17.5\,mag \citep{Saito2012}.

In this work, the molecular clouds are selected from the work of \citet{Rice2016}, 
who have created a catalogue of massive 
molecular clouds in the Galactic plane (13\degr $< ~l~<$ 348\degr) with a 
consistent dendrogram-based decomposition of the 
CO data from the $^{12}$CO CfA-Chile survey \citep{Dame2001}. The Rice et al. 
catalogue contains 1,064 massive clouds. 
247 of them are located inside the VVV survey footprint. All of them are 
located in the fourth quadrant of the Galactic plane.

For each molecular cloud, we select stars from the 2MASS and VVV catalogues in a square that centres at the 
the cloud with a side length of its size ($\sigma_r$), 
i.e. $|l-l_0| < \sigma_r$ and $|b-b_0| < \sigma_r$, where $l_0$ and 
$b_0$ are the Galactic coordinates of the cloud centre. We require that 
the stars must have detections in both the $J$ and \ks\ bands 
and have uncertainties in the two bands smaller than 0.1\,mag. To combine the 2MASS and VVV data, we first convert 
the VVV magnitudes to the 2MASS ones with the transformation 
equations from \citet{Soto2013}. 
The transform equations from \citet{Soto2013} are derived from de-reddened 
magnitudes and colours of stars. In this work we have ignored the 
effects of the variation of transformation coefficients caused by the dust extinction.
The impacts on the transformed magnitudes and colours are likely to be small.
For example, a disk RC that suffers an extinction value of 
\aks = 1\,mag, will result in uncertainty in the colour $(J-K_{\rm S})$ of 
about 0.06\,mag.
Then we merge the 
2MASS and VVV data by using the 2MASS data for \ks $<$ 12.5\,mag, 
and the VVV data for \ks $\ge$ 12.5\,mag. 
For some of the molecular clouds, there are only a few stars 
located in the defined boxes due to their small sizes.
We exclude the clouds that contain less than 5,000 stars. It yields 230 clouds in our sample. 
Fig.~\ref{catalog} shows the spatial distribution of these molecular clouds.
In the left panel of Fig.~\ref{method}, we show the ($J-$\ks, \ks) CMD of 
an example cloud. There are 108\,973 stars, with 3\,700 and 105\,273 stars from 2MASS 
(\ks $<$ 12.5\,mag) and VVV (\ks $\ge$ 12.5\,mag),
respectively.

\section{Method}

\begin{figure*}
    \centering
 \includegraphics[width=0.9\textwidth]{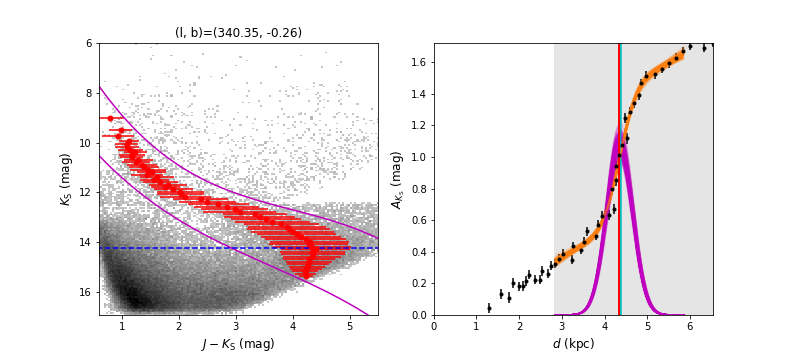}
  \caption{{\it Left panel}: CMD for an example molecular 
  cloud centred at ($l~,b$) = (340.35\degr, $-$0.26\degr). 
  The grey-scale image shows the number densities of the stars. 
  The magenta solid curves delimit the region where the RCs lie. 
  Filled circles with errorbars show locations of the RC peak density and 
  their 1$\sigma$ extent widths for the individual magnitude bins. Blue 
  dashed line shows the adopted cutoff magnitude.  {\it Right panel}: \ks-band extinction 
  profile for the sightline toward the example cloud (filled circles).
  Orange solid curves and magenta Gaussians are the fitted extinction profiles \aks$(d)$ and
  dust extinction models of the cloud $A^1_{K_{\rm S}}(d)$ of 
  100 randomly selected accepted models from the MCMC chain.
  Cyan lines denote the positions of the peaks of the individual Gaussian curves.
  The shadowed regions mark the distance priors we adopted for the fit
  and the red vertical line marks the resultant distance of the cloud $d_0$.}
  \label{method}
\end{figure*}

To obtain the extinction and distance profile of the sightline toward a given molecular cloud using the RCs, 
we first determine an empirical track of RCs by linking all peaks of the colour histograms of the RC candidates 
in the individual magnitude bins from the ($J-$\ks, ~\ks) CMD. 
Similar to the previous works (e.g., \citealt{Gao2009} and \citealt{Wang2020}), 
we check the CMD by naked eyes to achieve a rough stripe of RCs. 
The borders of the RC candidates are defined as two cubic polynomial 
function that fit seven points manually selected.  
We then select stars inside the boundaries and divide them into different horizontal bins 
according to their \ks\ magnitude. As the stellar 
densities differ from different \ks\ magnitude, we choose every 25 stars a 
bin for \ks$<$10\,mag and every 0.1\,mag a bin for \ks$>$10\,mag. 
For each \ks\ bin, we fit the distribution of the stellar colours 
with a Gaussian function to obtain its peak colour and 
the corresponding width. In the left panel of Fig.~\ref{method}, 
we show an example of how the RC track is defined. 
Due to the completeness limit of the observing data, the RC track 
moves to leftward for those faintest magnitude bins. 
We manually define a cutoff magnitude for each cloud to avoid the effect.

The resultant RC track is then used to measure the extinction profile of 
the sightline of the cloud. 
As RCs have almost a constant intrinsic color and absolute magnitude, 
we can easily convert their colours and magnitudes
($J-$\ks, ~\ks) into distances and extinction values ($d$,~\aks) by,
\begin{equation}
A_{K_{\rm S}} = 0.473 \times (J-K_{\rm S} - 0.7), 
\end{equation}
and 
\begin{equation}
d = 10^{0.2 \times (K_{\rm S}-(-1.61)-A_{K_{\rm S}})+1},
\end{equation}
where we adopt the extinction coefficient of 0.473 from \citet{Wang2019}, the 
intrinsic color of RC $(J-$\ks$)_0$ = 0.7\,mag from \citet{Grocholski2002} and the 
absolute magnitude of RC $M_{K_{\rm S}}$ = $-$1.61\,mag \citep{Alves_2000}. 

We finally determine the distance of a cloud based on the extinction profile of the cloud sightline. 
We fit the extinction profile within the distance range of the cloud by, 
\begin{equation}
A_{K_{\rm S}}(d)=A_{K_{\rm S}}^{0}(d)+A_{K_{\rm S}}^{1}(d),
\end{equation}
where $A_{K_{\rm S}}^{0}(d)$ is the \ks-band extinction contributed by the diffuse medium
and $A_{K_{\rm S}}^{1}(d)$ contributed by the dust within the molecular cloud at distance $d$. 
For the diffuse component, we have tested different models, such as the fourth-order polynomial \citep{Chen1998}, 
the quadratic polynomial (\citealt{Chen2017}, \citealt{Yu2019}, 
and \citealt{Zhao2020}), linear polynomial \citep{Marshall2009}, 
and constant (\citealt{Schlafly2014} and \citealt{Chen2020}), 
and decide to adopt the linear polynomial, as,
\begin{equation}
A_{K_{\rm S}}(d)= a*d+b,
\end{equation}
where $a$ and $b$ are polynomial coefficients.
Assuming a simple Gaussian distribution of dust in the cloud (\citealt{Chen2017}, \citealt{Yu2019}, and \citealt{Zhao2020}), 
the extinction profile of the cloud $E^1(d)$ is then given by, 
\begin{equation}
A_{K_{\rm S}}^{1}(d)=\frac{\delta A_{K_{\rm S}}}{2}{(1 + \rm erf}\left(\frac{d-d_0}{\sqrt{2} \delta d}\right)),
\end{equation}
where $\delta d$ is the width of the extinction jump, $\delta A_{K_{\rm S}}$ is the total extinction 
contributed by the dust grains in the cloud and  $d_0$ is the distance of the cloud.

For each cloud, the polynomial coefficients $a$ and $b$, the total extinction $\delta A_{K_{\rm S}}$, and the 
distance of the cloud $d_0$ are free parameters to fit. It is difficult for us to constrain the width of the extinction jump $\delta d$ 
in our model since there is significant degeneracy between $\delta d$ and $\delta A_{K_{\rm S}}$. We tried to fit $\delta d$
with several different priors, but those fits failed for a large number of clouds. Finally we decide to adopt a fixed $\delta d$ for each cloud. 
As discussed by \citet{Chen2020}, for distant clouds, the widths of the extinction jumps are mainly dominated by the
distance uncertainties of the extinction profile. In the current work, we assume the width of the extinction 
jump of each cloud consists of two components, the size of the cloud and 
the distance uncertainty, i.e., $\delta d = \sqrt{r^2 + \sigma_d^2}$, 
where $r$ is the size of the cloud ($r = d_0\sigma_r$), and $\sigma_d$ is the uncertainty of the distance. 
The typical distance uncertainty of RC is 7\,per\,cent (see Sect.~5.1). We thus adopt $\sigma_d = 0.07d_0$.
This assumption works well for our model, and we are able to successfully fit most of the extinction jumps.

A simple Bayesian scheme is adopted to obtain the free parameters and their statistical
uncertainties. We adopt the likelihood
\begin{equation}
L = \Pi \frac{1}{\sqrt{2 \pi \sigma^2}}\exp (\frac{-(A^{\rm obs}_{K_{\rm S}} (d) - A^{\rm mod}_{K_{\rm S}}(d))^2}{2 \sigma^2}),
\end{equation}
where $A^{\rm obs}_{K_{\rm S}}(d)$ and $A^{\rm mod}_{K_{\rm S}}(d)$ are the \ks-band extinction profile derived from the 
RC track and that modeled by Equation (2)--(4). 

For the distance $d_0$, we adopt a prior based on the kinematic distance of the cloud $d_{\rm kin}$ by,  
 \begin{equation}
P(d_0) =
\begin{cases}
1       & {\rm if~ no~}  d_{\rm kin} \\
1       & {\rm for~} d_{\rm kin} > 0 {\rm ~and~} |(d_0 - d_{\rm kin})/d_{\rm kin} | < 0.5 \\
1       &   {\rm for~} d_{\rm kin} > 0 {\rm ~and~} |d_0 - d_{\rm kin} | < 1.5 \\
0		&  {\rm otherwise},  \\
\end{cases}
\end{equation}
which means that we are only looking for the molecular clouds around its kinematic distance 
within 50\,per\,cent of its kinematic distance 
or 1.5\,kpc if the 50\,per\,cent kinematic distance is too small.
If there are near-far distance ambiguity, we search for the cloud around both the near and far kinematic distances.
Since the \citet{Rice2016} catalogue provides only one kinematic distance value for each cloud, either the near
of the far ones, we calculated the kinematic distances of all the selected clouds by the same 
algorithm as \citet{Rice2016} but with Galactic and Solar parameters from \citet{Reid2019}. 
Our derived kinematic distances are almost the same as those from \citet{Rice2016}.
If there is no kinematic distance resulted from the algorithm, we adopt a flat prior for all distances.
For other parameters, we adopt flat priors,
\begin{multline}
P(a, b, \delta A_{K_{\rm S}}) =   \begin{cases} 1 \quad \text{if} \,\,\begin{cases} 0 \,\, \leq \,\, a \,\, \leq  \,\, 0.3 \\
-0.8 \,\, \leq \,\, b \,\, \leq \,\, 1.0 \\
0.1 \,\, \leq \,\, \delta A_{K_{\rm S}} \,\, \leq \,\, 1.5 
 \end{cases}\\
 0 \quad \text{else}.
 \end{cases}
\end{multline}

The MCMC algorithm {\it emcee} \citep{Foreman2013} is applied in the current work.
The resulting distance and its statistical uncertainty are estimated in terms of the
50th, 16th and 84th percentile of the distance samples from the MCMC chain.
In the right panel of Fig.~\ref{method}, we show an example of how the extinction profile is fitted. 

\section{Results}

\begin{figure*}
    \centering
 \includegraphics[width=0.9\textwidth]{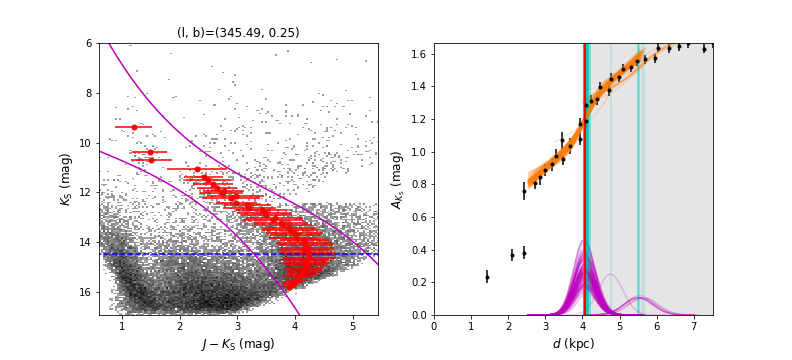}
 \includegraphics[width=0.9\textwidth]{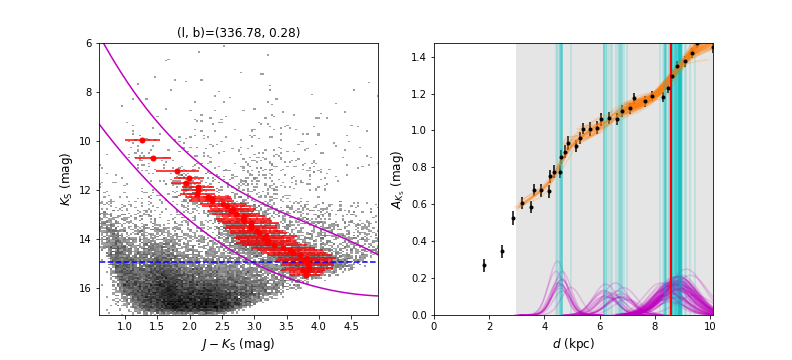}
 \includegraphics[width=0.9\textwidth]{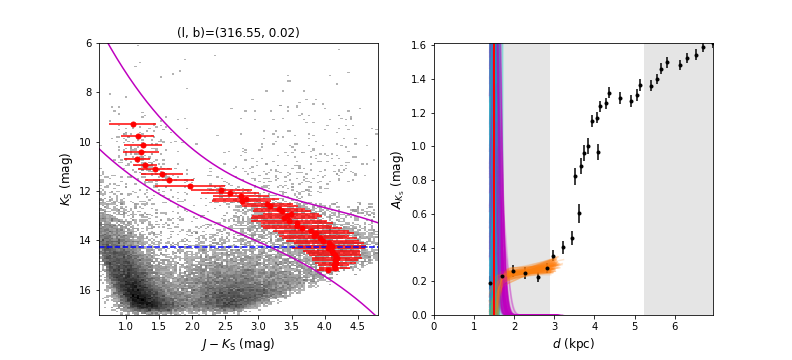}
  \caption{Distance determinations for three clouds centered at ($l,~b$) of (345.49\degr, 0.25\degr), 
  (336.78\degr, 0.28\degr), and (316.55\degr, 0.02\degr), respectively. See Fig.~\ref{method} caption for details.}
  \label{disest}
\end{figure*}

\begin{figure}
    \centering
  \includegraphics[width=0.42\textwidth]{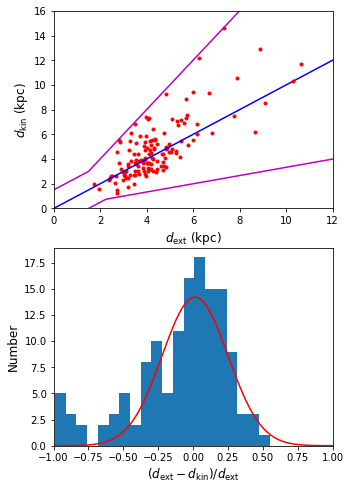}
  \caption{Comparison of our derived extinction distances of the clouds
  $d_{\rm ext}$ and their kinematic distances $d_{\rm kin}$. 
The magenta lines mark the boundaries of  our distance prior and the blue line 
denotes complete equality to guide the eyes. The red line in the bottom panel is
a Gaussian fit to the distribution of differences of values. }
  \label{dkincomp}
\end{figure}

\begin{figure}
    \centering
  \includegraphics[width=0.42\textwidth]{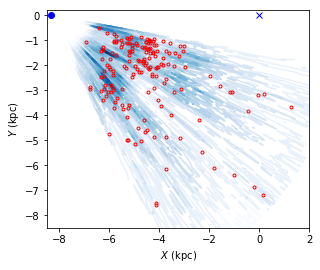}
  \caption{Spatial distribution of the molecular clouds catalogued in the current work (circles) in the X-Y plane, 
  plotted over the distribution of the Galactic dust of disk $|b|~<$ 2\degr\ (blue scales; \citealt{Marshall2006}). 
  The Sun and the Galactic Centre, assumed to be at the position of ($X,~Y$) = ($-$8.34, 0)\,kpc and 
    (0, 0)\,kpc, are marked as blue circle and cross, respectively.}
  \label{marshallcomp}
\end{figure}

\begin{table*}
    \centering
    \caption{Distances of molecular clouds in the fourth Galactic quadrant.}
    \begin{tabular}{rrrrrrrrr} % 5 columns, alignment for each
        \hline
          \hline
           $l$ & $b$ & $\sigma_r$ & $d_0$ &   $\delta$\aks\  & $R$ & $M$ & $V_{\rm LSR}$ & $d_{\rm R16}$ \\
          ($\degr$) & ($\degr$) & ($\degr$) & (kpc)  &  (mag) &  (pc) & (10$^3$\,M$_{\odot}$)  & (\kms)  & (kpc) \\
          \hline
  296.10 &    $-$0.22 &  0.20 &     5.29$\pm$0.69   &     0.11   &  35   &    39.42   &    13.47   &     8.40   \\
  297.83 &     0.70 &  0.42 &     3.35$\pm$0.31   &     0.12   &  46   &   190.70   &   $-$33.46   &     3.89   \\
  298.89 &    $-$0.43 &  0.28 &     3.26$\pm$0.33   &     0.11   &  30   &    98.52   &   $-$37.98   &     4.03   \\
  298.93 &    $-$0.16 &  0.12 &     8.66$\pm$1.22   &     0.14   &  34   &    88.78   &   $-$19.39   &     6.17   \\
  299.27 &     0.18 &  0.09 &     8.59$\pm$0.62   &     0.30   &  25   &    47.44   &    11.32   &     9.01   \\
  301.38 &    $-$0.27 &  0.12 &     5.86$\pm$0.85   &     0.13   &  23   &    65.34   &   $-$38.46   &     4.34   \\
  301.75 &     0.92 &  0.39 &     3.96$\pm$0.37   &     0.15   &  51   &   389.25   &   $-$41.00   &     4.39   \\
  301.79 &    $-$0.01 &  0.15 &     5.86$\pm$0.68   &     0.13   &  29   &    57.10   &    23.71   &    10.57   \\
  303.04 &     0.47 &  0.53 &     6.15$\pm$0.48   &     0.19   & 108   &  1325.45   &   $-$33.02   &     3.31   \\
  303.44 &     1.67 &  0.11 &     3.61$\pm$0.32   &     0.33   &  13   &         &   $-$37.60   &         \\
  304.04 &     1.02 &  0.13 &     3.91$\pm$0.39   &     0.19   &  16   &    45.84   &   $-$35.96   &     3.56   \\
  304.16 &     1.43 &  0.18 &     4.46$\pm$0.57   &     0.12   &  26   &    33.00   &   $-$45.09   &     4.68   \\
 \hline
    \end{tabular}
\parbox{\textwidth}{\footnotesize \baselineskip 3.8mm
The Table is available in its entirety in machine-readable form 
in the online version of this manuscript and also at the website \\
 ``http://paperdata.china-vo.org/diskec/rjcloud/table1.dat''.}
\end{table*}

We have applied the distance estimation algorithm to all the selected 230 giant molecular clouds. 
Similar as in the example shown in Fig.~\ref{method}, we can clearly see one extinction jump 
produced by the corresponding molecular cloud for a majority of our catalogued clouds. 
Our extinction models can nicely fit those extinction jumps. 

There are numerous high-density clouds in the Galactic plane, which
tend to overlap with each other along the sightlines. One can find more than one 
extinction jumps from the extinction profiles. In the top and middle panels of 
Fig.~\ref{disest}, we show two examples that our MCMC sampling procedure
finds more than one extinction jumps within the distance ranges of the corresponding cloud.
All those extinction jumps are nicely fitted by the accepted models of the MCMC analysis.

For the case of the cloud centred at ($l,~b$) of (345.49\degr, 0.25\degr) (top panel of Fig.~\ref{disest}), 
the extinction jump at $\sim$ 4\,kpc is much more significant than those at $\sim$4.8 and 5.6\,kpc. 
Since \citet{Rice2016} presented only the giant massive molecular clouds in their catalogue, 
it is reasonable for us to assume that the catalogued cloud has the largest total extinction values ($\delta$\aks) 
and thus have the most significant extinction jump. 
In such a case, we adopt the distance of the most significant jump as the distance of our catalogued cloud. 

However, for the case of the cloud centred at ($l,~b$) of (336.78\degr, 0.28\degr) (middle panel of Fig.~\ref{disest}),
there are also three clouds identified by our MCMC sampling procedure. 
The total extinction values $\delta$\aks\ of the three clouds are comparable to each others that
we are not able to isolate our interested molecular cloud from its foreground or background ones.
In the case that there is no cloud seen having the most significant extinction jump with much 
larger total extinction value $\delta$\aks\  than other identified clouds in the sightline, 
we fail to identify our catalogued cloud and will not provide its distance. 
Forty-one clouds in our sample are of this case.
 
Finally, for twenty clouds in our sample, we are not able to 
find any extinction jumps within the distance ranges around their kinematic distances. 
It is because that the kinematic distances of these clouds are either too near or too far away from us. 
Due to the magnitude limits of the 2MASS and VVV data ($\sim$10 -- $15$\,mag), 
we are only able to locate clouds 
that lie at distances between $\sim$2 and $\sim$11\,kpc from the Sun. 
In the bottom panel of Fig.~\ref{disest}, we show an example of such a case. 
The accepted extinction models of our MCMC sampling procedure 
are bad as there are no cloud seen within the fitting ranges. 
There is a cloud with a significant extinction jump at $\sim$4\,kpc, 
but its distance is not compatible to the kinematic distance of 
our interested cloud, which falls outside our fitting ranges. 
In such a case, we are not able to obtain the distance of the cloud.

As a result, we have successfully obtained distance estimates to 169  giant molecular clouds. 
The resultant extinction distances $d_0$ and their uncertainties 
and the total extinction values $\delta$\aks\ of these clouds are listed in Table~1. 
We also list their new radii $R$ and masses $M$ of the individual clouds estimated by our extinction distances, 
velocity $V_{\rm LSR}$ and kinematic distances $d_{\rm R16}$ from \citet{Rice2016} in the Table. 
These clouds range in distances from $d_0\approx2$\,kpc to $\sim 11$\,kpc.  
The uncertainties of the distance contain both the statistical uncertainties from the MCMC
analysis and the systematic uncertainties discussed in Sect.~5.1. 
Most of the clouds in our catalogue have statistical uncertainties smaller than 5\,per\,cent. 
For each of the molecular clouds, we have made figures analogous to Fig.~\ref{method}. 
They are available online\footnote{http://paperdata.china-vo.org/diskec/rjcloud/goodcloud.pdf}. 

In Fig.~\ref{dkincomp}, we compare the derived extinction distances $d_{\rm ext}$ to the 
kinematic distances $d_{\rm kin}$.  The kinematic distances $d_{\rm kin}$ are calculated by us with the 
rotation curve from \citet{Reid2019}. For some clouds, there are two kinematic distances.
We only adopt the one that is closer to the extinction distance in the case of distance ambiguity.    
Despite our adopted distance priors (Eq. 7), 
the newly established extinction distances deviate significantly
from their kinematic ones.
The differences between the extinction and the kinematic distances 
are a mean of 1.4\,per\,cent with an RMS scatter of 23.6\,per\,cent. 
Several clouds have extinction distances located near the lower limit of their kinematics distance ranges. 
There is an addition peak visible in the left of the 
distribution of the differences (bottom panel of Fig.~\ref{dkincomp}). 
It is partly due to the magnitude limits of the adopted photometric data in the current work that we are only able to trace the extinction distances 
within $\sim$11\,kpc from the Sun. 

In this work, we present one of the largest homogeneous catalogues 
of distant molecular clouds with the accurate direct measurement of distances. 
To verify the robustness of our distance estimates, we compare the spatial distribution of our molecular
clouds to the 3D dust distribution of the Galactic plane from \citet{Marshall2006}.
Based on the photometric data of 2MASS and the Besan\c{c}on Galactic model, 
\citet{Marshall2006} presented a 3D map of the inner Galaxy. 
The comparison is shown in Fig.~\ref{marshallcomp}. The spatial distribution
of the molecular clouds catalogued here is generally consistent
with the dust distribution of \citet{Marshall2006}. 
Most of the molecular clouds are located in the regions with high dust extinction,  
which validates the robustness of results in both papers.

\section{Discussion}

\begin{figure}
    \centering
  \includegraphics[width=0.42\textwidth]{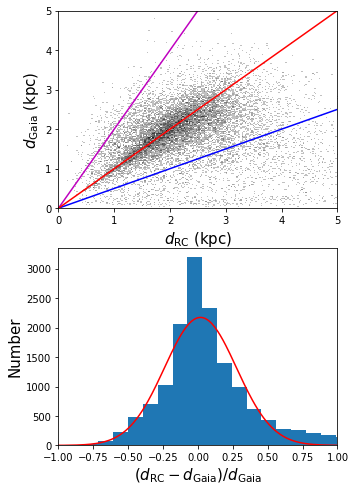}
  \caption{Comparison of the distances of the RC candidates derived from the RC assumption $d_{\rm RC}$ and 
those from Gaia DR2 parallaxes $d_{\rm Gaia}$.  The magenta, red and blue lines mark the 
$d_{\rm Gaia} = 2d_{\rm RC}$, $d_{\rm Gaia} = d_{\rm RC}$ and $d_{\rm Gaia} = d_{\rm RC}/2$, respectively.
The red line in the bottom panel is
a Gaussian fit to the distribution of differences of values.}
  \label{rccon}
\end{figure}

\subsection{Systematic uncertainties in cloud distance}

The statistical uncertainties of cloud distances from the MCMC sampling chains are mostly less than 5\,per\,cent. 
However the fitting errors are only parts of the final distance uncertainties of the clouds, 
which also include systematic uncertainties. 
The systematic uncertainties are intended to
account for several limitations, which we will discuss below.

In the first step of our method, we selected RC candidates that lie in the two borderlines. 
The selected candidates may contain some contaminants, such as the red giants or dwarfs. 
We cross-match the selected RC candidates of all the catalogued clouds with the Gaia DR2 data \citep{Gaia2018}. 
Two distances are simultaneously estimated for the individual RC candidates. 
One is Gaia distance $d_{\rm Gaia}$ obtained from the Gaia DR2 parallaxes \citep{Lindegren2018}, 
which can be treated as their `true' distances. 
The other is RC distance $d_{\rm RC}$ estimated by Eqs.~(1) and (2). 
If a selected candidate is not a RC, its RC distance $d_{\rm RC}$ 
will be over-estimated (for dwarfs) or underestimated (for giants) comparing to its Gaia distance $d_{\rm Gaia}$.  
In the current work, we adopt the Gaia distances of stars from \citet{Bailer2018} and exclude all 
sources with Gaia DR2 distance uncertainties larger than 20\,per\,cent.
The comparison of these two distances is shown in Fig.~\ref{rccon}. 
For most of the selected candidates, their RC distances are in good agreement with the Gaia ones. 
The mean and scatter of the differences are only 1.8 and 25.8\,per\,cent, respectively. 
The contamination of the giants or dwarfs does not affect the determination of the RC tracks, 
and it does not affect the distance uncertainties.

Our distance estimates are ultimately based on the distance estimates of RCs. 
The dispersions of the intrinsic colour $(J-$\ks$)_0$ and absolute magnitude $M_{K_{\rm S}}$ of the disk RCs 
are both $\sim$ 0.05\,mag \citep{Salaris2002, Chen2017b, Plevne2020}.  
Typically, for a point in the RC track located at $(J-$\ks$)_0$ = 2.5\,mag and \ks=13\,mag, 
the corresponding distance uncertainty is about $\sim$7\,per\,cent. 

The extinction coefficient adopted in the current work is from \citet{Wang2019}, 
who obtained high-precision optical to mid-IR extinction coefficients based on RCs selected from the APOGEE survey. 
The uncertainty of the near-IR extinction coefficient is about 5\,per\,cent (see also \citealt{Alonso2017}).
For $E(J - K_{\rm S})$ = 2 mag, the corresponding distance uncertainty is $\sim$2\,per\,cent.
Hence, to address the uncertainty of our resultant cloud distance, we adopt a systematic uncertainty of 7\,per\,cent. 

\begin{figure*}
    \centering
  \includegraphics[width=0.71\textwidth]{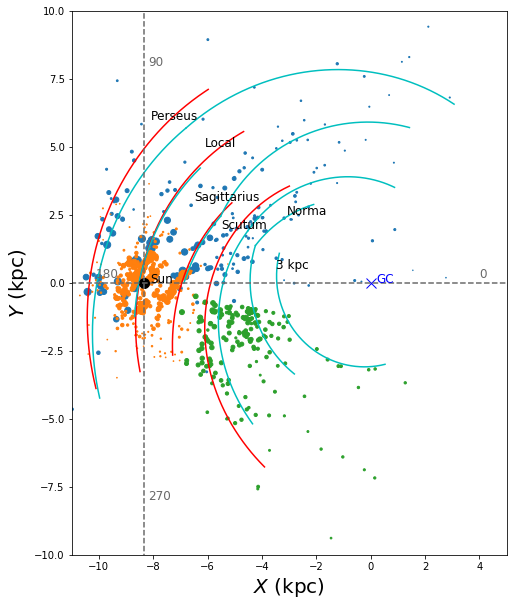}
  \caption{Distributions of the molecular clouds catalogued in the current 
  work (green dots) in the Galactic coordinates,
 complemented with the molecular clouds from 
 \citet[orange dots]{Chen2020} and HMSFRs from \citet[blue dots]{Reid2019}. 
The size of the dots indicates the inverse of the distance uncertainties.
The red and cyan solid curved lines denote the spiral arm models 
of, from left to right, the Perseus, Local, Sagittarius, 
Scutum-Centaurus, Norma and Near 3\,kpc arms, respectively
from \citet{Chen2019b} and \citet{Reid2019}. The Sun and the Galactic Centre, 
assumed to be at the position of ($X,~Y$) = ($-$8.34, 0)\,kpc and 
 (0, 0)\,kpc, are marked as black circle and blue cross in the plot, respectively.
 The directions of $l$ = 0\degr, 90\degr, 180\degr\ and 270\degr\ are also marked in the plot.}
  \label{xyplot}
\end{figure*}

\subsection{Spiral structure in the fourth Galactic quadrant}

The spiral structure of the Milky Way is not yet well determined,
especially for the fourth quadrant of the Galactic disk, due to the
lack of spiral tracers with accurately determined distances (e.g., \citealt[][]{Xu2018,Reid2019}).
Giant molecular cloud has proven to be one of the best tracers of Galaxy spiral arms (e.g., \citealt[][]{gcbt88,Xu2018}).
In our catalogue, the clouds are all giant molecular clouds having masses larger than 10$^4$\,M$_{\odot}$ (see Table~1). With
their robust determined distances in this work, the spiral
structure in the fourth Galactic quadrant may be better delineated.

In Fig.~\ref{xyplot}, we plot the projected distribution of the 169 giant molecular clouds in the
Galactic plane. These clouds are located primarily in the
Scutum-Centaurus Arm, the Normal Arm and the Near 3\,kpc Arm. Some of
the clouds seem to reside in the inter arm regions, which may be related with
arm branches or spurs. To make a comparison, we over-plot those
high-mass star-forming regions (HMSFRs) with trigonometric parallax
measurements as listed in \citet{Reid2019}  and those nearby molecule clouds with
extinction distances as listed in \citet{Chen2020}. Also shown in Fig.~\ref{xyplot} are
the best fitted spiral arm models from \citet{Reid2019} and \citet{Chen2019b}, which are updated views of the
Galaxy spiral structure. 

It is shown that the Scutum-Centaurus Arm
traced by our catalogued clouds deviates clearly from the best fitted model of \citet{Reid2019}.
The extensions of the Reid et al. spiral arms in the fourth
Galactic quadrant seem to be less reliable, as they were determined by
matching the observed spiral arm tangences rather than parallax
measurements. As shown in the Fig.~1 of \citet{Reid2019}, the HMSFRs
with parallax distances are still very rare in the fourth Galactic
quadrant. 

Recently, \citet{Chen2019b} estimated the parameters of
the Scutum-Centaurus Arm based on their large sample of O and early
B-type stars with the Gaia results of parallaxes, and also the HMSFR
sample of \citet{Xu2018} with trigonometric parallax
measurements. From Fig.~\ref{xyplot}, it is clear that comparing to the model of
the Scutum-Centaurus Arm from \citet{Reid2019}, the Chen et
al. model matches better to the distribution of our catalogued clouds. This is
reasonable, since there are more spiral tracers having accurate
distances in the fourth Galactic quadrant than that of \citet{Reid2019}. 

Similar situation is found for the Norma Arm, there are
also a large deviation between the best fitted model of \citet{Reid2019} and our 
giant molecular cloud distribution. 
Finally, eight of our catalogued
clouds are located near the 3\,kpc Arm of \citet{Reid2019}, which may
confirm the robustness of their model of the Near 3\,kpc Arm. To obtain
more accurate spiral structure in the fourth Galactic quadrant,
parallax measurements of more HMSFR masers are necessary. It would be
also helpful to combine together the data of HMSFR masers, giant molecular clouds, O and
early B-type stars in the analysis.

\section{Conclusion}

In this paper, we have estimated extinction distances to 169 
giant molecular clouds in the fourth Galactic quadrant. 
Based on the near-IR photometry of 2MASS and VVV, we have selected 
230 giant molecular clouds from \citet{Rice2016}. 
For each cloud, we select the RC candidates from the CMD of the cloud overlapping sightline. 
Based on the RC candidates, we find the RC track to determine the extinction profile of the sightline. 
Using a simple extinction model, we have derived accurate distance of the cloud by 
fitting the extinction and distance relation along the sightline. 
As a result, we have obtained extinction distances to 169 clouds, 
which locate at distances between $\sim$2 and $\sim$11\,kpc. 
The typical statistical error and the systematic uncertainty 
of the distances are $\sim$ 5 and 7\,pre\,cent, respectively. 

The result is a unique catalogue of distant
molecular clouds in the inner Galaxy with robust directly-measured distances. 
Based on this catalogue, we have tested different spiral arm models from the literature. 
We find large deviations between the spatial distribution of our 
giant molecular clouds and the Scutum-Centaurus 
and the Norma arm models from \citet{Reid2019} in the Galactic fourth quadrant. 
To obtain more accurate spiral structure in the region,
parallax measurements of more HMSFR masers are necessary. 

\section*{Acknowledgements}

This paper is published to commemorate the 60th 
anniversary of the Department of Astronomy, Beijing Normal University. 
We thank the anonymous referee for her/his useful comments.
This work is partially supported by National
Natural Science Foundation of China 11803029, U1531244, 
11533002, 11833006, 11988101, 11933011, 11833009 and U1731308
and Yunnan University grant No.~C176220100007.  
L.G.H thanks the support from the Youth
Innovation Promotion Association CAS.

This publication makes use of data products from the Two Micron All Sky Survey, 
which is a joint project of the University of Massachusetts and the Infrared 
Processing and Analysis Center/California Institute of Technology, funded by 
the National Aeronautics and Space Administration and the National Science Foundation.

This work is based on data products from VVV Survey observations made with 
the VISTA telescope at the ESO
Paranal Observatory under programme ID 179.B-2002.

\section*{Data availability}

The data underlying this article are available in the article and in its online supplementary material.

\bibliographystyle{mn2e}
\bibliography{ricecloud}

\label{lastpage}
\end{document}